# AUGMENTED REALITY USAGE FOR PROTOTYPING SPEED UP

J. Šťastný, D. Procházka, T. Koubek, J. Landa

## Abstract

ŠŤASTNÝ, J., PROCHÁZKA, D., KOUBEK, T., LANDA, J.: *Augmented reality usage for prototyping speed up.* Acta univ. agric. et silvic. Mendel. Brun., 2011, LIX, No. 2, pp. 353–360

The integral part of production process in many companies is prototyping. Although, these companies commonly have high quality visualization tools (large screen projections, virtual reality), prototyping was never abandoned. There is a number of reasons. The most important is the possibility of model observation from any angle without any physical constraints and its haptic feedback. The interactivity of model adjustments is important as well. The direct work with the model allows the designers to focus on the creative process more than work with a computer. There is still a problem with a difficult adjustability of the model. More significant changes demand completely new prototype or at least longer time for its realization.
The first part of the article describes our approach for solution of this problem by means of Augmented Reality. The merging of the real world model and digital objects allows streamline the work with the model and speed up the whole production phase significantly. The main advantage of augmented reality is the possibility of direct manipulation with the scene using a portable digital camera. Also adding digital objects into the scene could be done using identification markers placed on the surface of the model. Therefore it is not necessary to work with special input devices and lose the contact with the real world model. Adjustments are done directly on the model. The key problem of outlined solution is the ability of identification of an object within the camera picture and its replacement with the digital object. The second part of the article is focused especially on the identification of exact position and orientation of the marker within the picture. The identification marker is generalized into the triple of points which represents a general plane in space. There is discussed the space identification of these points and the description of representation of their position and orientation be means of transformation matrix. This matrix is used for rendering of the graphical objects (e. g. in OpenGL and Direct3D).

augmented reality, prototyping, pose estimation, transformation matrix

European manufacturers face strong competitive pressure which makes them speed up the development and production process constantly. One of the reasons is the market globalization. In many countries extremely low production costs are possible. The prize is often a violation of ethical and safety principles that is unacceptable in most European countries. One of the possibilities how to fight with such a kind of production is to maintain a technological lead. Therefore it is necessary to focus on methods which increase the development and production process efficiency.

An integral part of development of many products is their design. It is well known that the process includes initial design phase in a form of sketches which is usually followed by an electronic visualization phase of potential solutions (3D models development). The following step is usually prototyping. The gist of creation of even non-functional prototype is to obtain a clear image about its design and practically test its ergonomics. (Is it comfortable to hold the device? What is the field-of-view from the rear window of the car?).

Promising methods for streamlining the last mentioned phase are the virtual and augmented reality. The experiments with deployment of these visualization techniques have been ongoing for many years. Although it is possible to present a lot of partial successes with their deployment, their usage is still not common.

In the following part we discuss possible reasons of their difficult integration to the production process. Based on this analysis, we propose a new method for improvement of the prototyping process using augmented reality and we describe a solution of the key problem in detail – pose identification of an object.

## METHODS AND RESOURCES

Various advanced visualization techniques such as large and stereoscopic projections are a common part of the second design phase – development of 3D models or as a part of a decision-making process, presentation or marketing. In prototyping phase the virtual reality technology is a well established tool almost exclusively for testing of functional properties. We can also mention the significance of CAVE (stereoscopic projection surrounding the observer) which can be used to simulate the interior of e.g. car or plane. In these applications CAVE can exceed the prototype creation – there are no limitations in size and it is possible to simulate the functionality of the interior. The typical example is an article by Dmitriev *et al.*, 2004 where advanced visualization of a car interior is presented. Not entirely resolved problem is a complete replacement of the real prototype with a virtual one. Such a replacement will be possible after fulfilling the following criteria: The model has excellent visual quality and it is possible to manipulate with it naturally (including the haptic feedback).

The first requirement can be fulfilled in some detail. This aspect is discussed in Ong-Nee, 2004, p. 15–42 and Choi-Cheung, 2008. Much more significant problem is the natural manipulation with the object which includes the haptic feedback (see Ong-Nee, 2004, p. 43–64). Prototyping tools which use various forms of stereoscopic projection usually offer a limited functionality in comparison to the reality. Current technologies allow haptic feedback simulation (see Kortum, 2008), however the naturalness of these methods is often problematic (e.g. Johnson *et al.*, 2005).

During the product design phase it is crucial for a designer not to be limited by any hardware and software constraints. These constraints could have significant influence on the quality of his work. The poor result of various virtual tools does not have to be caused by limited functionality but more likely by psychological barrier of the designer. These cases are relatively common in many virtual reality applications (Carroll, 2005). Because of these reasons the physical model was never completely replaced. The question is whether the replacement is all possible. At least until a completely new generation of visualization techniques will be developed. The hope in this area comes with so-called physical holograms (Iwamoto *et al.*, 2008).

An interesting contribution for the area mentioned is the augmented reality. It represents a certain form of compromise. A physical model is not completely replaced but its main disadvantage – difficult extendability – is suppressed. Virtual objects could immediately extend the real model thus they increase the interactivity of the adjustments. There is almost no experience with such an augment reality deployment. Current industrial augmented reality applications are focused on the construction and maintenance of complex systems only. Details can be found in Ong-Lee, 2004, p. 237–383, Bottecchia *et al.*, 2010 and Bimber-Ramesh 2005. The reason is particularly in the hardware limitations. Immersive augmented reality is almost exclusively realized be means of the extensive hardware. These solutions are acceptable for manual activities but hardly acceptable for creative process. Current applications of virtual and augmented reality have the following common features:

The biggest problems are caused by using special hardware and software especially lack of the hardware maturity and psychological aspects with its use. In both cases the result is dissatisfaction with the deployment of the AR technology. In order to eliminate this problem required approach must not limit the user. In that case even the psychological barrier – to adopt the solution – will cease to exist.

The related problem lies in no mainstream solutions for discussed applications. Solutions must be usually implemented directly for the customer. It leads to higher costs. Therefore the companies are very careful with the deployment of these technologies. The problem will cease to exist after general solution is developed. Similar problems were e.g. in the beginnings of common stereoscopic projection.

Augmented reality applications are not necessarily connected with a special hardware. Also the basic principle of augmenting the scene is general enough to be implemented as a framework independent on a given problem. There are already experiments with such frameworks. The example could be the *ARToolKit* framework (described in Kato, 1999).

The basic application functionality lies in the localization of a marker in a given image and in pairing the marker with a specific virtual object. This step is called a registration. The registration is usually done by using an optical sensor – e.g. a digital camera. The issue of optical tracking can be divided into several sub-problems:

- Localization of the marker – the marker can be e.g. a Ping-Pong ball or a square shaped image attached to a surface.
- Computation of the marker position and orientation – recalculating marker position is a very important part of registration. Resolving this sub-problem will ensure the right orientation of an augmented object.

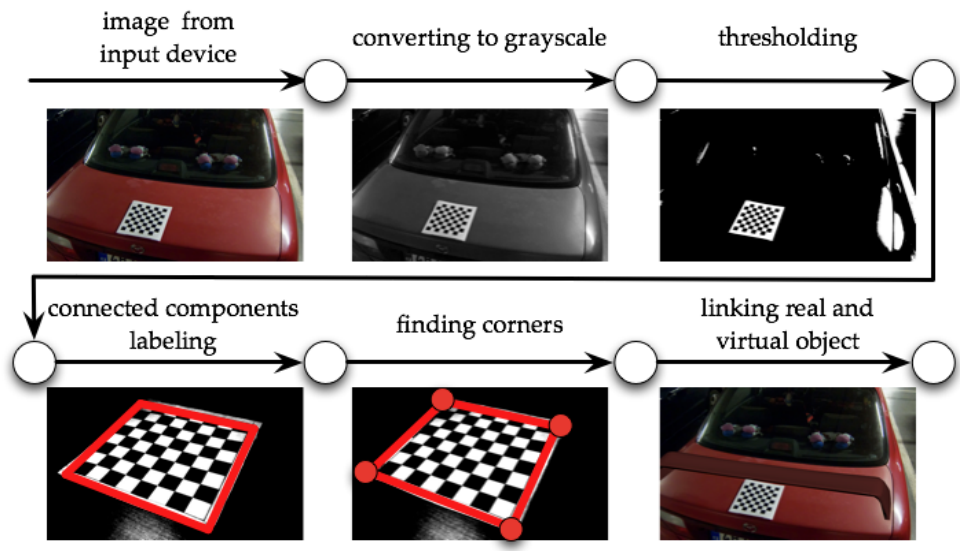

1: *General Solution for finding a marker*

General solution of the first sub-problem is presented in Fig. 1. In the first step an image which contains the marker is obtained. Localization of the marker could be done in several steps – image thresholding, connected components labelling and finally identification of the marker vertices. Then the localized marker must be projected to the camera plane and compared with a template stored in the application.

Before the object localization itself the image is usually preprocessed. That could reduce redundant information. The typically used method in preprocessing is thresholding. The scene is converted into a gray-scale and a threshold value must be determined. The threshold value is set according to the nature of the image. The pixel which has its value under the threshold is set to black and the pixel with the value above the threshold is set to white. This process is described in detail in Jähne, 2005. Thresholding the captured image allows to filter out a marker background. The result of this operation is the thresholded black and white image which is further processed.

The principle of connected components labelling lies in the comparison of neighbour pixel values. Each pixel is tested and in case of having the same value as a current component it is designated as a part of this connected component. Otherwise the pixel is labelled as the first pixel of a new component. More information about connected components is written in Acharya – Ray, 2005 or Di Stefano – Bulgarelli, 1999. After this labelling phase an image with highlighted individual components is obtained. These parts represent our potential markers.

The next step is localization of particular points of the marker. If the markers are e.g. squares it is the best to find their corners. The basic algorithm for finding corners is based on determining cornerness value for every pixel. This value determines the probability of every pixel being the corner pixel. From these points the cornerness map is created and the local maximums are found. There are different algorithms for finding out the cornerness value – e.g. Harris operator. The principle of the operator lies in comparing pixel values in small sectors of image. This algorithm is described e.g. in Mohanna – Mohktarian, 2001 or Rockett, 2003.

After performing all these operations the corners of the marker are available. For proper marker registration it is now necessary to create a transformation which allows us to project the object given by these corners from the camera perspective to its original shape. Then the comparison with the template could be done and in case of the positive result a virtual object could be inserted. The exact method of the projection must be defined by virtue the required functionality of the application. Our proposed method is described in the following section.

**RESULTS**

Our solution is based on the requirements outlined above. It represents a complete universal application for the augmented reality. It is possible to use it independently on a particular problem. Primarily it also does not expect the use of any special hardware which could limit the psychological acceptance of the solution. The basic idea is based on usage of well known approaches for identification of selected objects in an image. Our application for the augmented reality captures a given scene and identifies predefined objects be means of the digital camera. For the localized physical objects the application calculates their positions and orientations in space. Based on this information appropriate virtual objects are inserted into the image.

The proposed approach uses black and white squares in a shape of identification markers. Simi-lar markers are used e.g. in already mentioned *AR-ToolKit* framework (see Kato, 1999 and Popyrev, 2000). Their advantage is that they can be easily distinguished from the background of the image. The marker is printed on a hard surface and complemented by a mount that can be used to attach the marker onto the prototype. The size of the marker depends on the size of the prototype.

The operator imports virtual the 3D models into the AR application and pair them with their corresponding markers. During the work the designer has the set of markers which can be placed on the prototype. It is possible to change their position, orientation, etc. The gist is that the designer does not have to use a computer. The designer's creative process is not limited at all. It is only extended by the possibility to add a virtual object on a required place. An example of the use is testing the appearance of various rearview mirrors types on prototype of a car.

The related issue is the visualization of blending real and virtual objects. In the case of immersive AR application there are used glasses based on video image composition. However this type of glasses could significantly limit the nature of the work. Typical restricting factors are: the real world image digitalization (and related color distortion, lower resolution, etc.), the limited field-of-view and in many cases also the weight of equipment. We recommend the usage of this type of glasses only as an additional tool. The recommended primary tool for visualization of blending is a large screen projection. The large screen projection is used for similar purposes e.g. in Škoda Auto a.s. However the immersivity of such a solution is low. Nevertheless it does not limit a user in any way and allows to share the augmented image among many users.

The supplementary method of visualization lies in using a tablet. The tablet containing digital camera captures the scene and presents the image of prototype augmented by virtual objects. The tablet represents a certain mobile "window" into an augmented reality. This kind of the augmented reality is widely used for navigation. A very popular application is e.g *Layar* (http://www.layar.com/). The concept of *Layar* is described in Ebling-Caceres, 2010. The advantages of this solution are the ease of the use and higher immersivity in comparison to the static projection. However the visualization technique can be chosen freely and changed during the design process. The described application can be used with any of the discussed devices.

The only thing the application requires is a camera which will be used to capture the augmented prototype. The application reads images periodically from that camera. For each image the preprocessing and localization of a potential object are done (as described in previous section). These steps are followed by a marker registration. The marker registration process is implemented and described in further section.

## Pose identification

In the beginning we have a marker altered by a perspective projection and rotation in a space. This marker generally represents a plane (usually called an object plane). The second plane is the reference plane identical with the camera plane. The difference between these two planes defines the orientation of the marker in the scene. The plane could be represented for simplicity by a triple of points creating a triangle.

The found marker can be described, in a basic coordinate system, by a transformation matrix. This matrix can be perceived as a transformation from the basic coordinate system into the target coordinate system (the camera plane).

The issue with two pose identified triangles lies in finding appropriate reference and the measured location of an object. This is done by using two coordinate systems and with determination of relationship between these systems. Then the transformation between these two coordinate systems is done with the rotation around an imaginary axis and the adjustment of a position vector of the initial point of these systems.

The transformation algorithm can be described as follows:
- Choice of coordinate system for ideal object position (camera plane) and choice of coordinate system for recorded object position (object plane) based on the same characteristic points of triangle.
- Calculation of transformation matrix between those two coordinate systems.
- Recalculating this transformation to transformation of marker's end point.

The choice of the coordinate system of reference and real object positions is considered as crucial for the whole calculation. As shown in previous text, there is theoretically an infinite number of ways of choosing the coordinate system of an object. If the triangles were the same, it would not matter which point and vector are used for their pose identification. If they were not, different options could produce different deviations in pose identified objects. This is one of the main advantages of this process where we can choose among many solutions (without the need to modify the basic algorithm of calculation). We can also consider the suitability of different solution options for a specific case.

The choice of coordinate system of the given object can be described as follows:
- Choice of the initial point of coordinate system as one of the points of the triangle.
- Choice of the first axis of coordinate system as a vector between the beginning of this coordinate system and next points of the triangle.
- Choice of the second axis as the normals to the triangle surface.
- Choice of the third axis as the normals to the triangle surface given by the first and the second axis of coordinate system.

The position of the centre point of coordinate system does not have to correspond with the initial point of each axis. In determining coordinate systems it is just the matter of their directions. Obviously we choose only one point and one vector. We calculate the next two axes of coordinate system. If the first vector lies on the triangle surface the orthogonality of all vectors to each other is guaranteed at the same time. A very important condition for successful calculation is the need to determine coordinate systems with both triangles in the same way.

That is why when setting up coordinate systems it is appropriate to use clearly definable points of the triangle such as vertices, centres of gravity or centre points of individual sides.

Calculations for choosing the coordinate system are based on the following basic conditions: calculation of the vector defined by two points of the triangle, calculation of the normal vector to the triangle surface and calculation of the centre of gravity of the triangle (see Šťastný and Motyčka, 2009 or Škorpil and Šťastný, 2009).

The calculation of the transforming matrix of translation between two coordinate systems is given by the difference between individual parts of position vectors of these coordinate systems.

Generally it is important to think about further displacement which is created by rotating the object around end point of the marker. This displacement does not have to be taken into consideration when the center of the object rotation is placed in the initial point of coordinate system of the given object.

The calculation of rotation transformation matrix (between two coordinate systems) can be done in the following way: If A is an orientation matrix of coordinate system of the real object position, $R(\gamma)$ is a transformation matrix of rotation (by the angle $\gamma$) and B is a final orientation matrix, then $B = R(\gamma) \times A$.

This process can be used to get final orientation matrix of the given object from known orientation matrix and the transformation one. In order to get the transformation matrix it is necessary to multiply matrix B by the inverse matrix to matrix A. $R(\gamma) = B \times A^{-1}$.

The limiting factor is the invertibility of matrix A which does not have to be possible in case the determinant of this matrix is null. There are theoretically infinitely many ways how to connect the pose identified triangle with coordinate system. Because of that it is always possible to find the coordinate system whose orientation matrix is inverted without problems.

This transformation matrix $R(\gamma)$ can be incorporated into well-known rotation-translation matrix $\Theta$ which is widely used for projection of graphical objects.

$$\Theta = \begin{bmatrix} 0 & -r_z & r_y & t_x \\ r_z & 0 & -r_x & t_y \\ -r_y & r_x & 0 & t_z \\ 0 & 0 & 0 & 0 \end{bmatrix},$$

where $t_x, t_y, t_z$ represent a translation of a vertex and $r_x, r_y, r_z$ are normalized coefficients representing the rotation around appropriate axis. The translation is

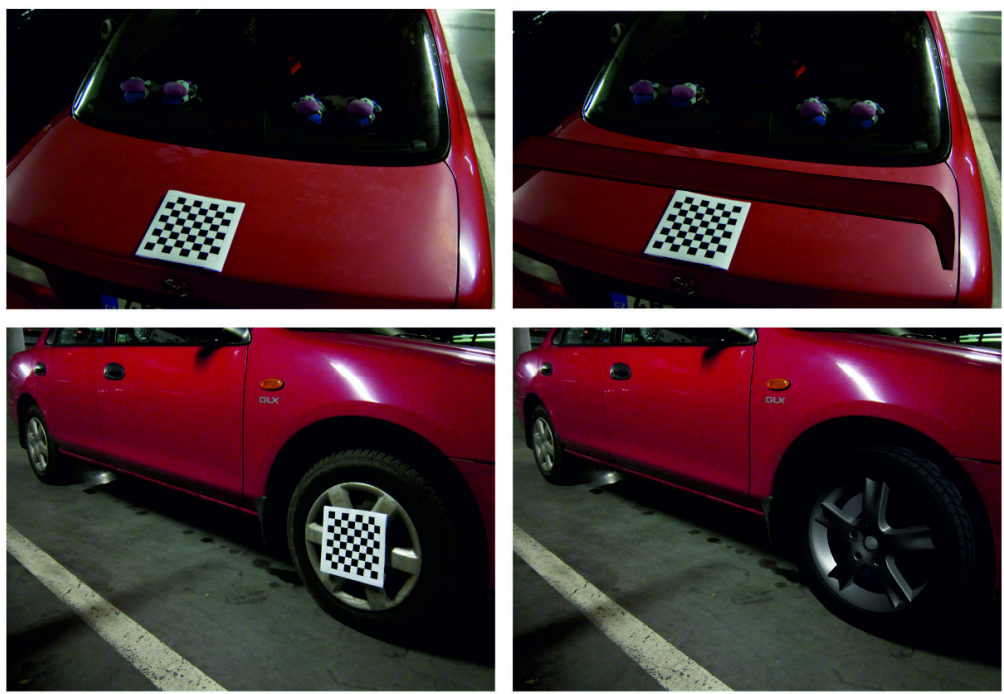

**Original scene**  **Augmented scene**

2: *Original scene with attached marker and scene augmented by a virtual object*

given by marker's position in the image and the rotation is given by previously defined matrix $R(\gamma)$.

The transformation matrix is in phase of rendering virtual object applied on vertices of the inserted object. Hence the object is placed into a required position and in a required angle. The application of transformation matrix $\Theta$ on the vertices of virtual object is done directly by the OpenGL library.

During the implementation arise a problem with high angles where the marker is extremely deformed by the perspective. In these situations the plane is lost and it is necessary to extrapolate the marker position and orientation. Without this extrapolation the simulation could have uncomfortable behaviour for the user. There is a number of approaches, from standard linear extrapolation based on few previous positions to many robust algorithms. Important limitation is the performance of the algorithm. Many robust methods could not be used in real-time applications. For the comparison of selected approaches see Škorpil and Šťastný, 2008 and Škorpil and Šťastný, 2006.

## DISCUSSION

A very interesting trend in augmented reality is so called Spatial Augmented Reality. It is a projection of virtual objects onto the real ones by using projectors. This technology is detailed in Bimber *et al.*, 2005 and its brief summary is in Bimber *et al.*, 2007. The augmented Scene is captured by a camera which allows the calibration of projection. It is even possible to project onto uneven and color surfaces. This type of AR cannot replace methods discussed earlier. In some cases inserted virtual objects are found outside of real object and therefore the projection on this object is not possible. However it could significantly contribute to increasing interactivity with the prototype.

## SUMMARY

The restraints of augmented and virtual reality deployment into the production process are especially the use of an expensive special hardware limiting the users and the absence of general, commonly available applications. These aspects lead to relatively high actual costs and high expectations related which are not fulfilled in many cases. The result is dissatisfaction with the deployment of these technologies.

Our proposed solution reflects these problems. The discussed application is capable of work both with a special hardware (glasses used for real and virtual image composition) and with widely used output devices such as the large screen projections and tablets. There is an example of product design described in our solution that does not restrain the creative process of a designer by using special hardware or software. The application allows to insert the virtual objects into a real scene according to the designer's requirements without the usage of a computer. The last part of the article is focused on the issue of object pose identification what is the crucial part of object registration – technology used for augmenting by the virtual objects. This description could be used for implementation of the problem in similar projects. However, proposed solution is not connected solely with the design process. It could significantly improve any process where the user is dealing with a complex data, from urban planning to bioimaging (see Martíšek *et al.*, 2007). Described optical tracking allows the user to control artificial models more naturally, than classical input interfaces.

Acknowledgements

This paper is written as a part of a solution of project IGA FBE MENDELU 14/2010 and research plan FBE MENDELU: MSM 6215648904.

Address

doc. RNDr. Ing. Jiří Šťastný, CSc., Ing. David Procházka, Ph.D., Ing. Tomáš Koubek, Ing. Jaromír Landa, Ústav informatiky, Mendelova univerzita v Brně, Zemědělská 1, 613 00 Brno, Česká republika, e-mail: jiri.stastny@mendelu.cz, david.prochazka@mendelu.cz, tomas.koubek@mendelu.cz, jaromir.landa@mendelu.cz